\documentclass[twocolumn,twoside,slac_two]{revtex4}
\usepackage{graphicx}
\usepackage{fancyhdr}
\newcommand{\lsim}{\mathrel{\rlap{\raisebox{.3ex}{$<$}}
    \raisebox{-.6ex}{$\sim$}}}

\begin{document}

\title{The CLEAR Experiment}

%

\author{K. Scholberg
}
\affiliation{Department of Physics, Duke University, Durham, NC 27708, USA}
\author{T. Wongjirad}
\affiliation{Department of Physics, Duke University, Durham, NC 27708, USA}
\author{E. Hungerford}
\affiliation{Department of Physics, U. of Houston, Houston, TX 77204, USA}
\author{A. Empl}
\affiliation{Department of Physics, U. of Houston, Houston, TX 77204, USA}
\author{D. Markoff}
\affiliation{North Carolina Central University, Durham, NC 27695, USA}
\author{P. Mueller}
\affiliation{Physics Division, Oak Ridge National Laboratory, TN 37831, USA}
\author{Y. Efremenko}
\affiliation{Department of Physics and Astronomy, U. of Tennessee, Knoxville, TN 37996, USA}
\author{D. McKinsey}
\affiliation{Department of Physics, Yale University, New Haven, CT 06520, USA}
\author{J. Nikkel}
\affiliation{Department of Physics, Yale University, New Haven, CT 06520, USA}

\begin{abstract}
The Spallation Neutron Source in Oak Ridge, Tennessee, is designed to
produce intense pulsed neutrons for various science and engineering
applications.  Copious neutrinos are a free by-product.  When it
reaches full power, the SNS will be the world's brightest
source of neutrinos in the few tens of MeV range.  The proposed CLEAR
(Coherent Low Energy A (Nuclear) Recoils) experiment will measure
coherent elastic neutral current neutrino-nucleus scattering at the
SNS.  The physics reach includes tests of the Standard Model.
\end{abstract}

\maketitle

\thispagestyle{fancy}


\section{Neutrino Production at the SNS}

The Spallation Neutron Source (SNS) is a recently-completed facility located at Oak Ridge National Laboratory, TN: it provides the most intense pulsed neutron beams in the world for use in a wide range of science and engineering studies.   The beam is pulsed at 60 Hz and the expected power in the first phase is 1.4 MW.  First beam was attained in 2006, and the power has been gradually increasing.  Full power is expected in 2010.  Some upgrades are envisioned for the next decade, including a power upgrade to 2-5 MW, and possibly a second target station.

\begin{figure}[!htbp]
\centering
\includegraphics[height=2.1in]{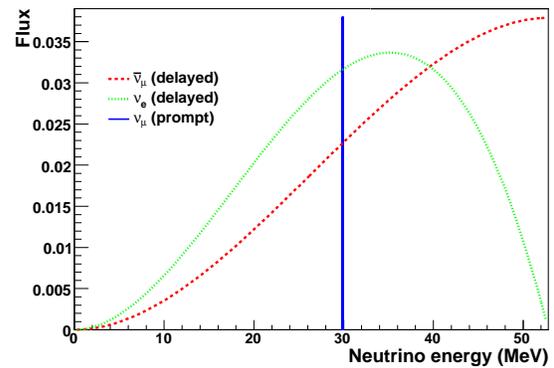}
\hspace{0.8in}
\caption{Stopped-pion neutrino spectrum, showing the different flavor components.}
\label{fig:sns_spec}
\end{figure}

\begin{figure}[!htbp]
\centering
\includegraphics[height=1.95in]{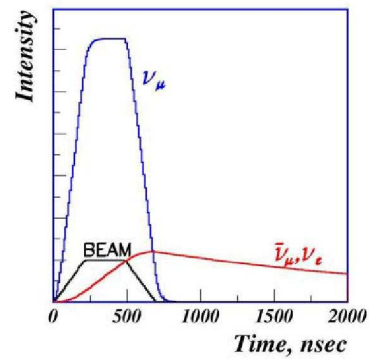}
\caption{Timing of the SNS pulse with respect to the neutrino fluxes.}
\label{fig:sns_timing}
\end{figure}

Neutrinos are produced as a free by-product when protons hit the SNS
target.  The collisions produce hadronic showers including pions.
Whereas $\pi^-$ get captured, $\pi^+$ slow and decay at rest.  The
$\pi^+ \rightarrow \mu^+ + \nu_\mu$ decay at rest produces a prompt,
monochromatic 29.9~MeV $\nu_\mu$.  The $\mu^+$ then decays on a 2.2
$\mu$s timescale to produce a $\bar{\nu}_\mu$ and a $\nu_e$ with
energies between 0 and $m_\mu/2$. The $\bar{\nu}_e$ flavor is nearly
absent from the flux.  See Figures~\ref{fig:sns_spec} and
\ref{fig:sns_timing}.  About 0.13 neutrinos per flavor per proton are
expected, which amounts to about 10$^7$ per flavor at 20~m from the
target~(\cite{Avignone:2003ep}).  The short-pulse time structure of
the SNS is also advantageous: for a 60 Hz rate, the background
rejection factor is a few times $10^{-4}$.

Past experiments have successfully used similar stopped-pion 
$\nu$ sources:
examples are LANSCE at LANL, which hosted
the LSND experiment~(\cite{Athanassopoulos:1996ds}), and ISIS at RAL, 
which hosted KARMEN~(\cite{Zeitnitz:1994kz}).    However the SNS has far superior characteristics for neutrino experiments compared to any existing or near-future source.

\section{Coherent Elastic Neutral Current Neutrino-Nucleus Scattering}

Coherent elastic neutral current neutrino-nucleus
scattering~(\cite{Freedman:1977xn}) has
never been observed.  In this process, a neutrino of
any flavor scatters off a nucleus at low momentum transfer $Q$ such
that the nucleon wavefunction amplitudes are in phase and add
coherently.  
The cross-section for a spin-zero nucleus, neglecting
radiative corrections, is given by~(\cite{Horowitz:2003cz}),

\begin{equation}
\left(\frac{d\sigma}{dE}\right)_{\nu A} = \frac{G_F^2}{2\pi} 
\frac{Q_w^2}{4} F^2(2ME) M \left[2 - \frac{M E}{k^2}\right], 
\end{equation}
where $k$ is the incident neutrino energy, $E$ is the nuclear recoil
energy, $M$ is the nuclear mass, $F$ is the ground state elastic form
factor, $Q_w$ is the weak nuclear charge, and $G_F$ is the Fermi
constant.   The condition for coherence requires that 
$Q\lsim \frac{1}{R}$,
where $R$ is the nuclear radius. 
This condition is largely satisfied for neutrino energies up to 
$\sim$50~MeV
for medium $A$ nuclei.
Typical values of the total
coherent elastic cross-section are in the range $\sim
10^{-39}$~cm$^2$, which is relatively high compared to other neutrino
interactions in this energy range ($e.g.$ charged current
inverse $\beta$ decay on protons has
a cross-section $\sigma_{\bar{\nu}_e p}\sim 10^{-40}$~cm$^2$, and 
elastic
neutrino-electron scattering has a cross-section\
$\sigma_{\nu_e e}\sim 10^{-43}$~cm$^2$).

In spite of its large cross-section, coherent elastic $\nu$A
scattering has been difficult to observe due to the very small
resulting nuclear recoil energies.  The maximum recoil energy is 
$\sim 2 k^2/M$, which is in the sub-MeV range for $k~\sim~50$~MeV and for
typical detector materials (carbon, oxygen).  Such energies are below
the detection thresholds of most conventional neutrino
detectors. 
However, in recent years there has been a surge of progress in the
development of novel ultra-low threshold detectors, many with the aim
of WIMP recoil detection or pp solar neutrino detection.  Thresholds
of $\sim$10~keV or even lower may be
possible.  Some of these new technologies, for instance noble liquids
~(\cite{McKinsey:2004rk}),
may plausibly attain ton-scale masses in the relatively near future.

\begin{figure}[!htbp]
\centering
\includegraphics[height=2.1in]{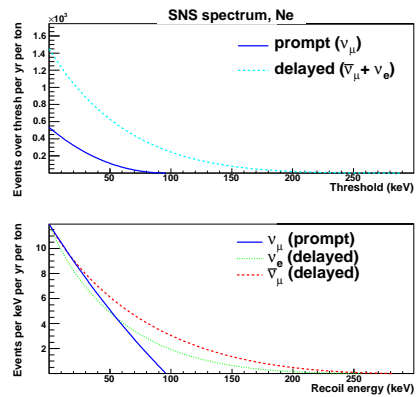}

\caption{Bottom: Differential yield at the SNS in one ton of 
Ne
(solid: $\nu_\mu$, dotted: $\nu_e$, dashed: $\bar{\nu}_\mu$) per year 
per keV,
as a function of recoil energy,
for one year of running at the SNS at 46 m from the target.
Top: Number of interactions over recoil energy threshold
(solid: $\nu_\mu$, dashed: sum of $\nu_e$ and $\bar{\nu}_\mu$),
as a function of recoil energy threshold. }
\label{fig:snsyield1}
\end{figure}

\begin{figure}[!htbp]
\centering
\includegraphics[height=2.1in]{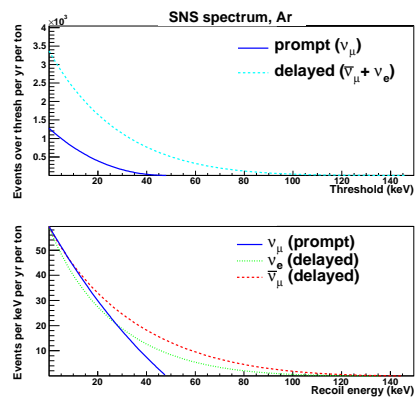}
\caption{Same as Figure~\ref{fig:snsyield1} for Ar.}
\label{fig:snsyield2}
\end{figure}

Although ongoing efforts to observe coherent $\nu A$ scattering 
at reactors~(\cite{Barbeau:2007qi,Collar:2008zz,Wong:2005vg})
are
promising, a stopped-pion beam has several advantages with
respect to the reactor experiments. Higher recoil energies bring
detection within reach of the current generation of low threshold
detectors which are scalable to relatively large target
masses. Furthermore, 
the pulsed nature of the source (see Figure~\ref{fig:sns_timing}) allows
both background reduction and precise characterization of the
remaining background by measurement during the beam-off period.
Finally, the different flavor content of the SNS flux means 
that 
physics sensitivity is complementary.
The expected rates for the SNS are quite promising for noble liquids
~(\cite{Scholberg:2005qs}): see Figures~\ref{fig:snsyield1}
and \ref{fig:snsyield2}.  

Coherent elastic $\nu$A scattering
reactions are important in stellar core collapse 
processes~(\cite{Freedman:1977xn}),
as well as being useful for core collapse supernova 
neutrino
detection~(\cite{Horowitz:2003cz}).  A rate measurement will
have bearing on supernova neutrino physics.
The $\nu$A coherent elastic scattering cross-section is
predicted by the Standard Model (SM), and form factor
uncertainties are small~(\cite{Horowitz:2003cz}).  Therefore a measured
deviation from prediction could be a signature of new physics.
Some possibilities are described below.

The SM predicts a coherent elastic scattering rate proportional to
$Q_w^2$, the weak charge given by $Q_w = N-(1-4\sin^2 \theta_W)Z$,
where $Z$ is the number of protons, $N$ is the number of neutrons and
$\theta_W$ is the weak mixing angle.  Therefore the weak mixing angle
can be extracted from the measured absolute cross-section, at a
typical $Q$ value of 0.04~GeV/$c$.  
A deviation from the SM prediction
could indicate new physics.  If the absolute cross-section can be
measured to 10\%, there will be an uncertainty on $\sin^2 \theta_W$ of
$\sim 5 \%$.  One might improve this uncertainty by looking at ratios
of rates in targets with different $N$ and $Z$, to
cancel common flux uncertainties; future use of enriched neon is a possibility.
There are existing precision
measurements from atomic parity
violation~(\cite{Bennett:1999pd,Eidelman:2004wy}), SLAC
E158~(\cite{Anthony:2005pm}) and NuTeV~(\cite{Zeller:2001hh}).
However there is no previous
neutrino scattering measurement in this region of $Q$.
This $Q$ value is relatively
close to that of the proposed Qweak parity-violating
electron scattering experiment at JLAB~(\cite{vanOers:2007if}).
However coherent elastic $\nu$A scattering 
tests the SM in a different channel and therefore
is complementary: we note that this is a first-generation experiment.

In particular, one can search for non-standard interactions (NSI)
of neutrinos with nuclei.  Existing and planned precision measurements
of the weak mixing angle at low $Q$ do not constrain new physics which
is specific to neutrino-nucleon interactions.
The signature of NSI is a deviation from the expected 
cross-section~\cite{Barranco:2005yy}.
Reference~(\cite{Scholberg:2005qs}) explores the sensitivity of an
experiment at the SNS.
As shown in the reference, under reasonable assumptions, if the rate
predicted by the SM is observed, neutrino scattering
limits more stringent than
current ones
~(\cite{Dorenbosch:1986tb, Davidson:2003ha}) 
by about an order of magnitude can be obtained.
Reference~(\cite{Barranco:2007tz}) looks at the sensitivity of a
coherent
$\nu$A scattering experiment to some specific
physics beyond the Standard Model, including models with extra 
neutral gauge
bosons, leptoquarks and R-parity breaking interactions.

Searches for NSI are based on precise knowledge of the nuclear form factors, which are known to better than 5\%~(\cite{Horowitz:2003cz}), so that a deviation from the SM prediction would indicate
physics beyond the SM.  
If we assume that the Standard Model is a good description,  then with sufficient
precision one can measure neutron form factors. 
(Reference~(\cite{Amanik:2007ce}) explores this possibility, which could
be within reach of a next-generation experiment.)
If a small deviation
from the SM prediction were to be observed, presumably one would have
to pursue additional measurements to determine whether the discrepancy
were due to nuclear physics or beyond-the-SM physics. It would
be interesting in either case.

\section{The CLEAR Experiment}

The
specific detector we plan to build is called CLEAR (Coherent Low
Energy A (Nuclear) Recoils).  We have selected a single-phase design which allows
interchangeable noble liquid target materials.  Multiple targets are desirable to test
for physics beyond the Standard Model. 

\begin{figure}[!ht]
  \centering
    \includegraphics[height=2.0in]{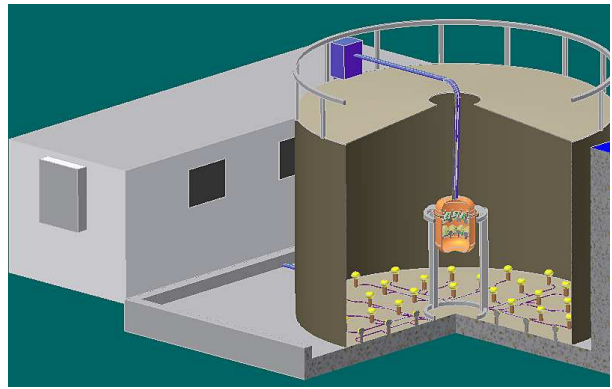}
  \caption{CLEAR experiment concept.  The cryogenic inner
    detector enclosed in a vacuum vessel will be positioned inside a tank
    of water, which provides neutron shielding and an active muon veto
    by detection of Cherenkov radiation with an array of PMTs.
   }\label{fig:overview}
\end{figure}

The CLEAR experiment at the SNS comprises an inner noble liquid detector
placed inside a water tank. The water tank will be instrumented with
photomultiplier tubes (PMTs) to act as a cosmic ray veto. An overview diagram
of the experiment is shown in Figure~\ref{fig:overview}.

\subsection{The Inner Noble Liquid Detector}

\begin{figure}[!ht]

  \centering
    \includegraphics[height=2.0in]{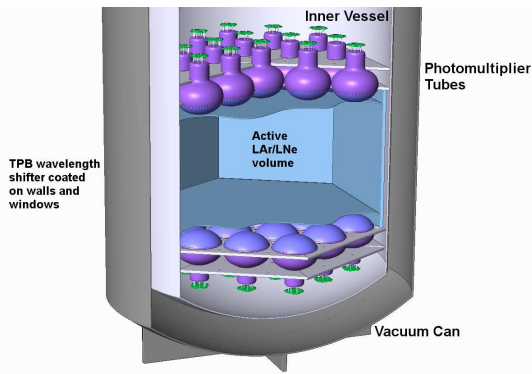}
  \caption{The inner detector, containing an active target of LAr or 
   LNe viewed by photomultipliers, as described in the text.
   }\label{fig:inner_det} 

\end{figure}

We propose to use liquid argon (LAr) and liquid neon (LNe) as the
detector materials for the first-generation CLEAR experiment.  LAr and
LNe are bright scintillators, comparable in light yield to NaI but
with a faster response. Several properties of LAr and LNe make this overall
approach attractive.  First,
LAr and LNe scintillate strongly in the vacuum ultraviolet and
are transparent to their own scintillation light, allowing for event
detection with a low energy threshold. LAr and LNe are dense
enough (1.4 and 1.2 g $\rm cm^{-3}$, respectively) to allow
significant target mass in a modest detector volume.  Pulse
shape discrimination (PSD) 
to select nuclear recoils is possible because both LAr and LNe have
two distinct mechanisms for the emission of scintillation light. 
These two scintillation channels, resulting from singlet molecule
decay and triplet molecule decay, have very different fluorescence
lifetimes and are populated differently for electron recoils than for
nuclear recoils.  This allows nuclear recoils and electron recoils to
be distinguished on an event-by-event basis.  This approach to
electron recoil discrimination has been proposed
for liquid neon~(\cite{McKinsey:2004rk}), and 
for liquid argon~(\cite{Boulay:2006}).  Demonstrations of 
discrimination in the energy window of interest have been
accomplished in the MicroCLEAN~(\cite{Nikkel:2008}), DEAP-I~(\cite{Boulay:2009zw}), 
and WARP~(\cite{Brunetti:2005}) experiments.
The ability to exchange LAr with LNe, with different sensitivities to
coherent neutrino scattering and fast neutrons, would allow both event
populations to be distinguished and characterized.  
Finally, argon and neon are relatively inexpensive detector materials.  

The CLEAR detector will be a cylindrical LAr/LNe scintillation
detector, with an active LAr (LNe) mass of 456 (391)~kg.  
The active volume will be about 60~cm in diameter, and 44~cm tall.
A schematic of the active detector is shown in
Figure~\ref{fig:inner_det}.  The central active mass will be viewed by
38 Hamamatsu R5912-02MOD photomulipliers (PMTs) divided into two
arrays, one on the top of the active volume
facing down, and the second array on the bottom 
facing up.  All PMTs will be completely immersed in the
cryogenic liquid.  A cylinder of PTFE will define the outer radius of the
active volume. The bottom and top of the active volume will be defined by 
two fused silica or acrylic plates.  
Ionizing radiation
events in the liquid cryogen will cause
scintillation in the vacuum ultraviolet (80~nm in LNe
or 125~nm in LAr), which is too short to pass through 
the PMT glass.  
The inner surface of the PTFE walls and end plates will 
be coated with a thin film 
of tetraphenyl butadiene (TPB) wavelength shifter.
The ultraviolet scintillation light is
absorbed by the wavelength shifter and re-emitted at
a wavelength of 440~nm. The photon-to-photon conversion 
efficiency is about 100\% for LAr scintillation
and about 130\% for LNe scintillation~(\cite{McKinsey:1997}) . The wavelength-shifted light is then detected
by the PMTs.   Using the MicroCLEAN detector we have verified that the chosen PMT model can be used immersed in
LAr or LNe.

The detector will be contained in a stainless steel vacuum cryostat. 
A pulse-tube refrigerator, mounted near the water tank, provides
cooling power to maintain the active fluid at the desired temperature
value.  The noble gas is continuously circulated, boiled,
purified and re-liquefied during operation to maintain a sufficiently
large light yield and triplet molecule lifetime.  Molecular impurities
that affect light collection are removed using gas-phase
recirculation through a commercial heated getter.  We have found this
approach to be highly effective in MicroCLEAN, and it
is the same approach used in the XENON~(\cite{Angle:2008we}) and LUX~(\cite{LUX}) experiments.

We will calibrate the detector using low-activity neutron sources such
as $^{252}$Cf or Am/Be to determine the nuclear recoil response, 
and gamma ray sources such as $^{57}$Co and $^{137}$Cs to determine the electronic
recoil response.  Calibration sources will be
introduced into the tank from above, inserted down a fixed stainless
steel tube until they are adjacent to the cryostat at a known
position.

\subsection{Siting and Shielding}

\begin{figure}[!ht]
\centering
\includegraphics[height=2.0in, angle=90]{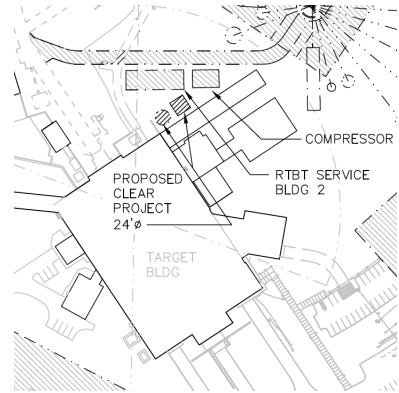}
       \caption{CLEAR site at the SNS.
}\label{fig:clear_site}
\end{figure}

The CLEAR experiment will occupy a site 46~m from the SNS target,
behind the beam (see Figure~\ref{fig:clear_site}).  

The detector will sit inside a standard steel agricultural water tank
of about 24~ft in diameter and 16~ft in height, which will serve both
for shielding and as a cosmic ray veto.  Additional steel shielding
blocks will also be employed.  The water tank will be instrumented
with 32 8-inch PMTs.  The inside of the tank will be protected against
corrosion and lined with Tyvek to increase reflectivity and light
collection, as was done for both Super-K~(\cite{Fukuda:2002uc}) and
KamLAND~(\cite{Eguchi:2002dm}) outer detectors.  Our
Geant4~(\cite{Agostinelli:2002hh,Allison:2006ve}) simulations show that
excellent cosmic veto efficiency is obtained with at least 20~PMTs,
and a configuration in which all PMTs are placed on the bottom of the
tank is near-optimal.

\subsection{Backgrounds}

We assume 
the SNS is
running at its full 1.4 MW power, and a live running time of
2.4$\times10^7$ s/yr for each of LAr and LNe.  The SNS is expected to be
running at full power by early 2010, which is before the anticipated
start time of CLEAR. With a nuclear recoil energy window between
20-120~keV (30-160~keV) and a 456 (391) kg LAr (LNe) target, we will
have about 890 (340) signal events from the muon decay flux, and about
210 (110) signal events from the prompt $\nu_\mu$ flux.  Backgrounds for
the $\nu$ signal detection come from neutrons (cosmic and SNS-related) and
misidentified gammas.

\begin{figure}[!htbp]

\centering
\includegraphics[height=2.1in]{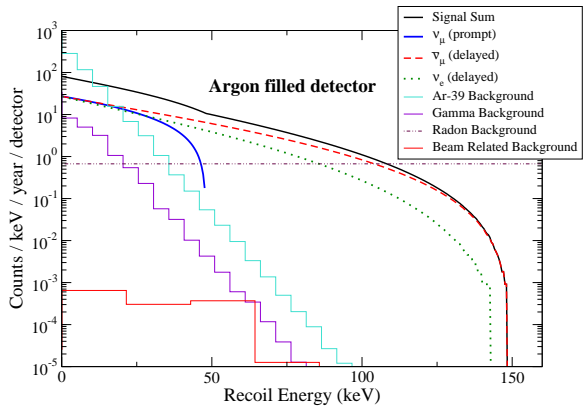}
\caption{Number of events in both neutrino signals along with beam
and detector related backgrounds, for a LAr filled detector (456~kg).}
\label{fig:Ar_sig_bg}
\end{figure}

\begin{figure}[!htbp]
\centering
\includegraphics[height=2.1in]{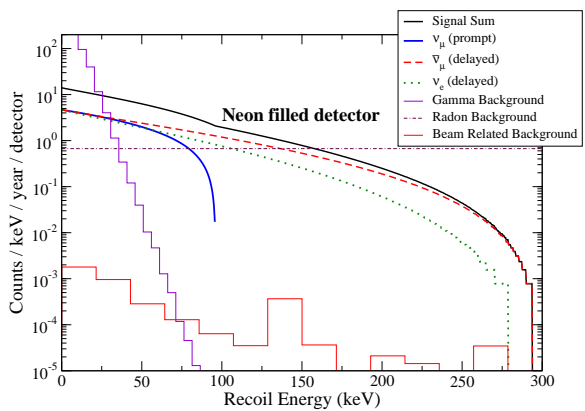}
\caption{Same as Figure~\ref{fig:Ar_sig_bg} for a LNe filled detector (391 kg).}
\label{fig:Ne_sig_bg}
\end{figure}

We divide the backgrounds into two main categories: beam-related and
non-beam-related.  Inelastic neutrino interaction backgrounds are
assumed negligible; not only are cross-sections an order of
magnitude smaller than the coherent elastic $\nu$A cross-section, 
but such events will produce electrons or gammas and
hence will be rejected by PSD selection. We have adopted several
strategies for estimating background.  For beam-related background, we
make use of simulations done by the SNS neutronics group
(\cite{nusns}, updated December 2007), and further propagate these
through the shielding using FLUKA 2008.3b.1 code~(\cite{Fluka1:2005, Battistoni:2007zzb}).  The background event rate is
estimated using a Geant4-based inner detector simulation. 
 For cosmic
ray-related background, we employed a cosmic-ray generator~(\cite{CRY})
and an independent Geant4 simulation of the water veto tank.
The inner detector
simulation code was also employed to estimate backgrounds due to
radioactivity of the detector materials.  These background estimates
are discussed in more detail below, and results are summarized in
Figures~\ref{fig:Ar_sig_bg} and \ref{fig:Ne_sig_bg}.

\noindent
\textbf{Beam-related backgrounds:} Beam-related neutrons, which can
cause nuclear recoils in the energy region of interest, are of concern
because if they occur within the 10 $\rm \mu$s beam timing window they
will be indistinguishable from signal on an event-by-event basis.  At
SNS beam turn-on in 2006 we began measurements of neutron and gamma
fluxes at a site inside the target building using 5 inch liquid
scintillator detectors with pulse shape discrimination capabilities. 
One of these was also used for measurements outside the target building.
The measured neutron flux showed little obvious correlation with beam 
power at the inside site through 2006.
These measurements give only a general order of magnitude estimate.
We expect changes as the nearby instruments turn on and beam power
changes.  However, we did observe that the fast neutron flux dropped
rapidly with respect to the beam, and was unobservable by our detectors
after a few $\mu$s.

\begin{figure}[!htbp]
\centering
\includegraphics[height=2.1in]{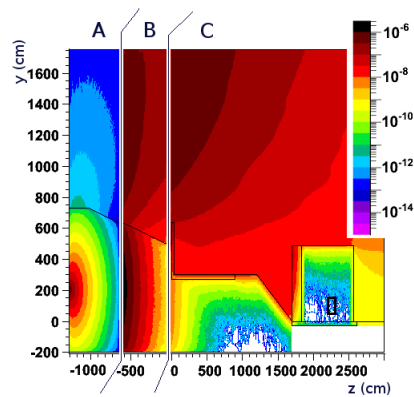}

 \caption{Predicted spatial distribution of neutrons given 
   as a fluence graph (tracklength density, cm/cm$^3$/primary).
   The color scale applies to stage C. Shown is a projection
   averaging over $\pm$ 150~cm with respect to the plane of interest
   which is perpendicular to the beam line and passes through the center
   of the CLEAR detector volume.}

\label{fig:fluka}
\end{figure}

\begin{figure}[!htbp]
\centering
\includegraphics[height=2.1in]{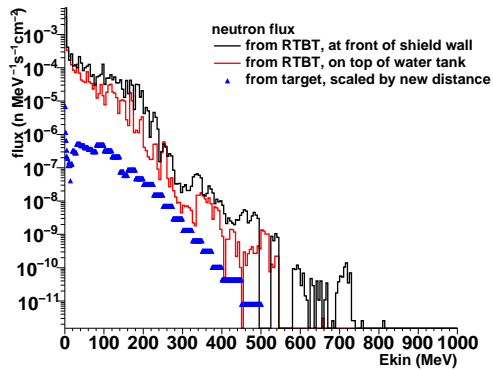}

 \caption{Predicted neutron flux just in
   front of the iron shield wall (black line) and also for neutrons
   entering the water tank from the top (red line). For comparison we
   include the predicted neutron background originating from the SNS
   target scaled to the CLEAR location. }

\label{fig:fluka2}
\end{figure}

The dominant beam-related background is neutrons produced from a beam
loss of 1 W/m in the transport line. The neutron flux from the
SNS target is much less than the contribution from the beam line, but has
the same spectral shape.  Neutrons from the instrumentation also have
the same spectral shape, but are highly dependent on the position of
the instruments, and cannot be properly evaluated at this time. 
The beam line is enclosed in
an approximately square tunnel with walls 75~cm thick and buried under
at least 550~cm of soil.  The neutronics group at the
SNS has simulated the neutron spectrum directly above
beamline and the soil, and at a perpendicular distance of 34~m from
the CLEAR detector.    
We implement in FLUKA a basic 3D model of
the site and a radial line source aligned with the beam line.
In order to maintain statistical accuracy at the higher neutron
energies, we throw neutrons in a flat kinetic energy distribution up to
1 GeV. Later the neutron energy distribution is imposed on the
results.
The propagation is accomplished in three stages (as shown in 
Figure~\ref{fig:fluka}) to speed the simulation, and to allow normalization to the
neutronics simulation mentioned above. 
The first stage, A, simulated the beam tunnel and
soil 
for a beam intensity of approximately
1~MW.   Figure~\ref{fig:fluka2} shows flux in the front of the steel
wall and on top of the water tank.
 The neutron flux into the target volume, as shown in stage C,
 is a projection onto the plane of the figure from -150~cm
 behind to 150~cm in front of the plane. The flat neutron spectrum has
 been filtered by the initial neutron spectrum.
 The figure shows that almost all of the background comes from
 ``sky shine'' and not directly from the beamline.  Although not shown,
 neutrons which leave the beamline with energies below
 approximately 200~MeV are absorbed in the shielding. Obviously, neutrons with
 higher energies degrade to lower energies, but with less
 intensity.  The neutrons (and gammas which are not shown) entering the target volume 
are then passed to the inner detector Geant4 simulation. 

\vspace{0.1in}

\vspace{0.1in}

These beam-related background numbers are our best estimate according to
currently known parameters. 
While they are comfortably small, they
still have considerable uncertainties.
Furthermore, characteristics of the background are expected to change 
as nearby instruments turn on and off.  For this reason, we plan comprehensive
background measurements onsite before detector installation and during
running.  
Based on the pre-installation measurements, we will
optimize the shielding configuration.  The detector stand will be movable to
allow for repositioning of
the inner detector within the water tank to allow for minimization of backgrounds.
The concrete pad will be designed with driven pilings,
so it can accommodate extra shielding if needed.
The
shielding itself will be available for very low cost in the form of
Duratek steel blocks.

\noindent
\textbf{Non-beam-related backgrounds:} These are mitigated by the SNS
beam window: the timing, shown in Figure~\ref{fig:sns_spec}, allows a
factor of $6 \times 10^{-4}$ rejection of steady-state backgrounds
assuming a 10 $\rm \mu s$ timing window.  Timing of individual events
in the detector can be known to within $\sim$10 ns using the fast
scintillation signal.  Furthermore, these backgrounds can be
subtracted using beam-off data.
Cosmic ray-related background can be
vetoed with high efficiency by our muon veto phototubes, and ambient
radioactivity will be reduced significantly by the water shielding. 
The estimated
rates of cosmic-ray neutrons
creating signal candidates during the beam window
are small
and are not shown in Figures~\ref{fig:Ar_sig_bg} and ~\ref{fig:Ne_sig_bg}.

We rely on pulse shape discrimination to
reduce $\gamma$ backgrounds.  At energies above a few tens of keV, LAr
and LNe scintillation detectors are capable of very good
discrimination, as described in
references~\cite{Nikkel:2008} and~\cite{Lippincott:2008}.  Gamma
backgrounds are dominated by the $^{238}$U, $^{232}$Th, and $^{40}$K
that are present in the PMT glass.  The contribution due to the
stainless steel cryostat and other materials is small in comparison. 
$^{39}$Ar, which is present in atmospheric argon at approximately 1
part in $10^{15}$, decays at a rate of about 0.8 Bq/kg of argon. 
While this is a relatively high rate, PSD is highly efficient at
removing this background, improving exponentially with energy. 
Figure~\ref{fig:Ar_sig_bg} shows that $^{39}$Ar background is smaller
than the signal above a nuclear recoil energy threshold of 20 keV,
given a beam timing cut of $6 \times 10^{-4}$.  Measurements of the
$^{39}$Ar rate without beam timing cuts will allow this background to
be measured accurately and statistically subtracted.  We note that it
may be possible to reduce $^{39}$Ar background significantly by
employing depleted argon from underground sources~(\cite{Galbiati:2007xz}).
In LNe, electron
recoil background is dominated by gamma ray Compton scattering.  This
occurs at a rate much less than the $^{39}$Ar rate in LAr, but the PSD
is less effective in LNe, resulting in a higher analysis
threshold of 30 keV.

Radon daughters are another background of concern, which may also be
substantially removed using beam timing cuts.  In particular,
$^{210}$Po, which has a 138-day half-life, can produce nuclear recoils
in the active volume that can mimic the signal from $\nu$A
scattering.  The inside surface of the active region is approximately
2.9~$\rm m^{2}$ in area.  By mechanically scrubbing the PTFE and fused
silica surfaces before TPB deposition, bagging these pieces in
radon-impermeable plastic during the time between deposition and
installation, and maintaining these surfaces in a HEPA-filtered
atmosphere during final assembly, we expect to be able to keep the
radon daughter decay rate below 100 m$^{-2}$\,day$^{-1}$ in the energy
region of interest.  This target value for CLEAR exceeds the radon
daughter background rate per unit area already demonstrated in the
DEAP-I experiment at SNOLAB~(\cite{boulayidm}), which contains similar TPB-coated acrylic surfaces.
The initial surface treatment and TPB coating of the PTFE and fused
silica will take place offsite, and final installation will
occur in a HEPA-filtered enclosure adjacent to the experimental site
at the SNS. After 
timing cut, radon daughter background corresponds to a background of about 100
events in the  year after installation.  As in the case of $^{39}$Ar
background, radon daughter background can be quantified without beam timing to
gain an accurate measurement of its rate.

\begin{figure}[!htbp]
\centering
\vspace{0.2in}
\includegraphics[height=2.1in]{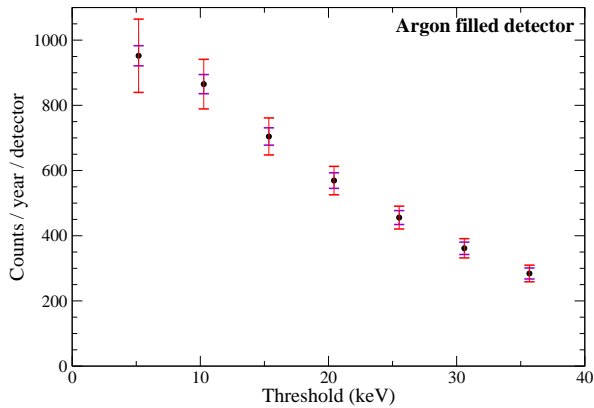}
\caption{Plots of the integrated signals as a function of analysis threshold,
for a LAr filled detector (456 kg).
The smaller error bars include just the statistical uncertainties on the
signal.  The larger error bars incorporate all systematic uncertainties including
those due to the background subtraction.  The total measured signal is suppressed by a factor 
of two in these plots due to a flat pulse shape discrimination cut.  In the real 
experiment an energy-dependent cut will preserve more of the signal at higher
energies.}
\label{fig:Ar_sig_thresh}
\end{figure}

\begin{figure}[!htbp]
\centering
\vspace{0.2in}
\includegraphics[height=2.1in]{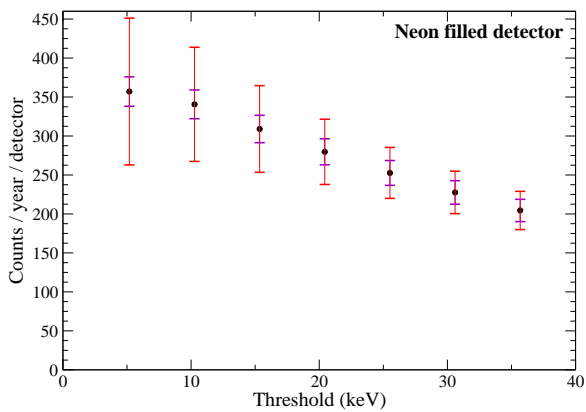}
\caption{
Same as Fig~\ref{fig:Ne_sig_thresh} for a
LNe filled detector (391 kg).}
\label{fig:Ne_sig_thresh}
\end{figure}

\subsection{Signal and Background Summary}

Figures~\ref{fig:Ar_sig_bg} and \ref{fig:Ne_sig_bg}
show the results of the background simulations as compared to the
expected signal for both argon and neon.  Plotted along with the
signal curves (and their sum) are the beam-related backgrounds, the
$^{39}$Ar backgrounds (for the argon case), an estimate of the radon
background, and the gamma background in both LAr and LNe.  We note
here also that the signal will have a characteristic time-structure (a
prompt component with known delay from the SNS target, and a $\mu$
decay component of known time constant), distinct from that of the
background.   After
folding in the above backgrounds and the 50\% detector efficiency, we
can calculate the expected total integrated signal as a function of
analysis threshold.  This is shown in Figures~\ref{fig:Ar_sig_thresh}
and~\ref{fig:Ne_sig_thresh}.

\noindent
\textbf{Systematic uncertainties:}
Assuming one
year of data, the statistical uncertainty on the rate measurement is estimated to be 4\% (8)\% for LAr (LNe).  The uncertainty in the neutrino flux is estimated to be
10\% and is currently the dominant uncertainty in this measurement.
The energy threshold uncertainty also contributes uncertainty to the
overall $\nu$A scattering cross-section.  The analysis threshold is
much above the trigger threshold, so the uncertainty in the energy
threshold is dominated by uncertainty in the energy scale.  The energy
scale will be determined through calibration using $\gamma$-ray sources
in combination with the known LAr and LNe nuclear recoil scintillation
efficiencies.  Overall we project a systematic uncertainty from energy
threshold to be 4\% (3\%) from LAr (LNe).  There is comparatively little
uncertainty in the target mass, as there are no fiducial cuts assumed
in the analysis, the density of the liquid is well known, and the
active volume can be measured accurately during assembly.  We estimate
the fiducial mass to be calculable with better than 1\% uncertainty.  Background uncertainties are in the few percent range.  Overall expected uncertainty on the coherent $\nu$A rate measurement is 12\% (13\%) for LAr (LNe).

\section{Summary}

The SNS creates an intense neutrino source in the few tens of MeV
energy range.  The CLEAR experiment aims to measure coherent elastic
$\nu$-A scattering by siting a single-phase noble liquid scintillation
detector 46~m from the SNS neutrino source.  The planned active target
mass is 456~kg of LAr or 391~kg of LNe.  Non-beam-related backgrounds
include cosmic rays, internal and external radioactivity, radon, and
$^{39}$Ar (for argon); all of these may be well characterized using
data outside of the beam window.  Beam-related neutron backgrounds,
which cannot be rejected using the beam time window, have been shown
using extensive simulations to be comfortably small.  The absolute
rate can be measured with $\sim$12-13\% uncertainty.

\begin{acknowledgments}
The CLEAR collaboration is grateful to the SNS neutronics group
for background simulations.

\end{acknowledgments}

\bigskip 
\bibliography{main_bib}



\end{document}